\documentclass[10pt, conference]{IEEEtran}
\usepackage[utf8]{inputenc}
\usepackage[detect-all]{siunitx}
\usepackage{array,tabularx,calc}
\usepackage{mathtools}
\usepackage{amsfonts}
\usepackage{graphicx}
\usepackage{capt-of}
\usepackage{amssymb}
\usepackage{amsfonts}
\usepackage{pbox}
\usepackage{textcomp}
\usepackage[]{algorithm, algpseudocode}
\usepackage{etoolbox}
\usepackage{url}
\usepackage{tikz}

\DeclareSIUnit{\belmilliwatt}{Bm}
\DeclareSIUnit{\dBm}{\deci\belmilliwatt}
\DeclareSIUnit{\belisotropic}{Bi}
\DeclareSIUnit{\dBm}{\deci\belisotropic}
\DeclareSIUnit{\bit}{bit}
\DeclareSIUnit{\byte}{B}

\usepackage[capitalise]{cleveref} 

\ifCLASSOPTIONcompsoc	\usepackage[caption=false,font=normalsize,labelfont=sf,textfont=sf]{subfig}
\else					\usepackage[caption=false,font=footnotesize]{subfig}
\fi

\makeatletter
\renewcommand{\ALG@beginalgorithmic}{\small}

\makeatother

\usepackage[noadjust]{cite}

\graphicspath{{Results/}}

\begin{document}

\usetikzlibrary{arrows}
\usetikzlibrary{shapes}
\newcommand{\mymk}[1]{%
	\tikz[baseline=(char.base)]\node[anchor=south west, draw,rectangle, rounded corners, inner sep=0.1pt, minimum size=3.5mm,
	text height=2mm](char){\ensuremath{#1}} ;}

\newcommand*\circled[1]{\tikz[baseline=(char.base)]{
	\node[shape=circle,draw,inner sep=0.1pt] (char) {#1};}}

\title{A Fast Gateway Placement Algorithm for Flying Networks} 

\author
{
	\IEEEauthorblockN{Gonçalo Santos, João Martins, André Coelho, Helder Fontes, Manuel Ricardo, Rui Campos}
	\IEEEauthorblockA{INESC TEC and Faculdade de Engenharia, Universidade do Porto, Portugal\\
	goncalo.r.santos@fe.up.pt, joao.miguel.martins@fe.up.pt, andre.f.coelho@inesctec.pt,\\ helder.m.fontes@inesctec.pt, manuel.ricardo@inesctec.pt, rui.l.campos@inesctec.pt}
}%

\maketitle

\begin{abstract}
The ability to operate anywhere, anytime, as well as their capability to hover and carry cargo on board make Unmanned Aerial Vehicles (UAVs) suitable platforms to act as Flying Gateways (FGWs) to the Internet. The problem is the optimal placement of the FGWs within the flying network, such that the Quality of Service (QoS) offered is maximized. The literature has been focused on optimizing the placement of the Flying Access Points (FAPs), which establish high-capacity small cells to serve the users on the ground, overlooking the backhaul network design, including the FGW placement. The FGW placement problem is exacerbated in highly dynamic flying networks, where the dynamic traffic demand and the movements of the users may induce frequent changes in the placement of the FAPs. 

The main contribution of this paper is a fast gateway placement (F-GWP) algorithm for flying networks that determines the optimal position of a FGW. With F-GWP, backhaul communications paths with high enough capacity are established between the FAPs and the FGW, in order to accommodate the traffic demand of the users on the ground. Simulation and experimental results show F-GWP is two orders of magnitude faster than its state of the art counterpart, while ensuring the same flying network performance.

\end{abstract}

\section{Introduction}
\begin{figure}
	\centering
	\includegraphics[width=1\linewidth]{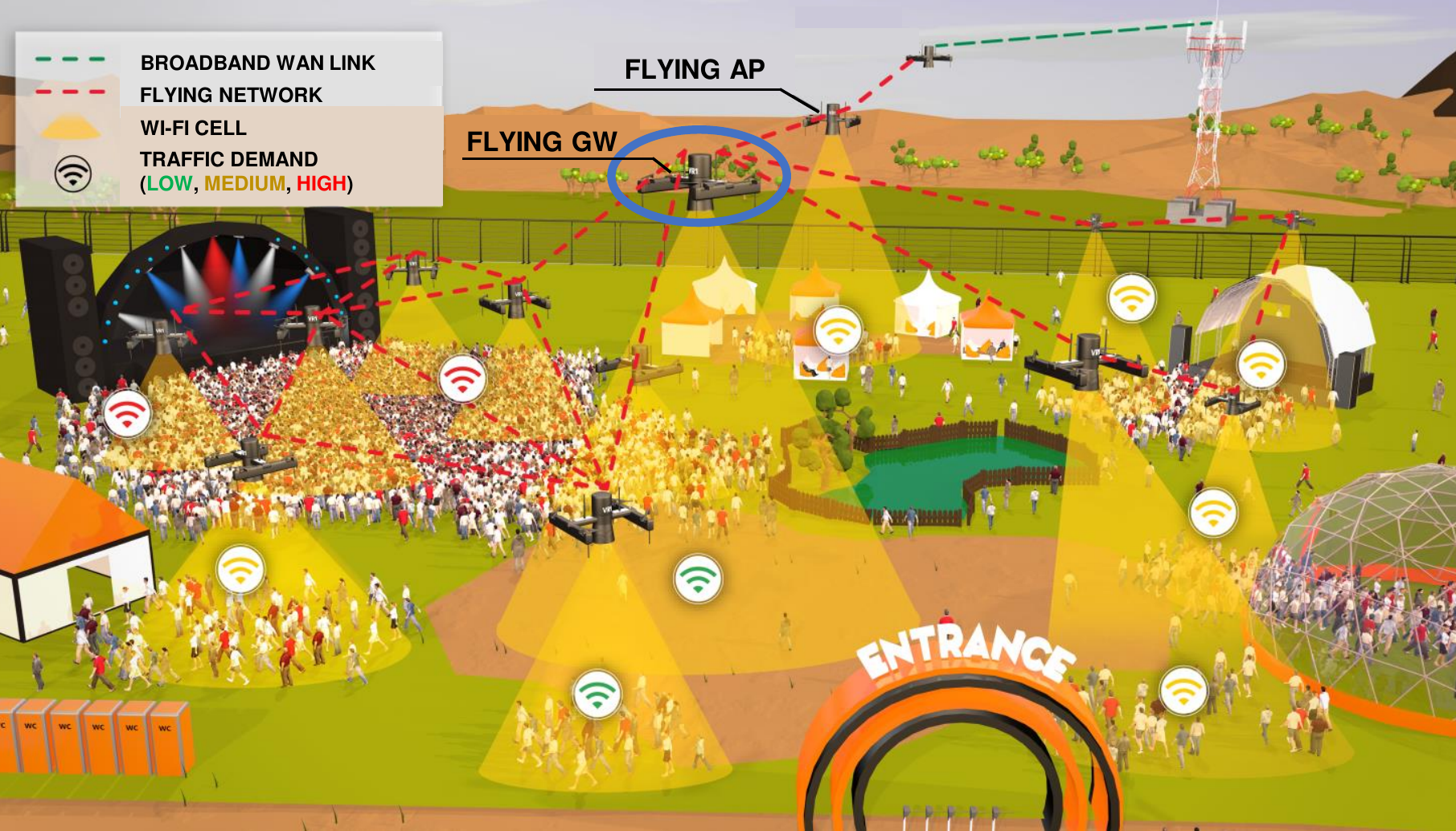}
	\caption{Flying network providing Internet connectivity to the users in a music festival \cite{WISE:online}.}
	\label{fig:music_festival_scenario}
\end{figure}

The ability to move in the three-dimensional space, hover above the ground, and carry cargo on board makes Unmanned Aerial Vehicles (UAVs) suitable platforms to act as communications nodes, such as Wi-Fi Access Points (APs) and cellular Base Stations (BSs) \cite{Zeng2016}. This is paving the way to the deployment of flying networks to provide temporary on-demand broadband wireless connectivity and to reinforce the capacity of existing networks in a myriad of scenarios, including disaster scenarios \cite{Zhao2019} and outdoor festivities \cite{Almeida2018}, as depicted in \cref{fig:music_festival_scenario}.\looseness=-1

A challenge inherent to flying networks is the optimal placement of the UAVs such that the traffic demand of the users on the ground and their Quality of Service (QoS) requirements are fulfilled. Several UAV placement solutions have been proposed in the literature \cite{Almeida2018,WISE:online,Mozaffari2015,Kalantari2017,Zeng2016_ThroughputMaximization,alzenad2017, lyu2016, Alzenad2018,yu2018,mozaffari2016_EfficientDeployment,chen2018,na2019}, but they are focused on enhancing the radio coverage and improving the QoS offered by the radio access network, which is formed by Flying Access Points (FAPs). Although the users on the ground are directly affected by the access network, the QoS offered by the latter is influenced by the backhaul network. Thus, the backhaul network needs to be carefully designed. A critical aspect in the backhaul network is the gateway placement.\looseness=-1

In \cite{Coelho2019}, a centralized traffic-aware gateway placement (GWP) algorithm for flying networks with controlled topology was proposed. GWP takes into account the knowledge about the offered traffic and the future positions of the FAPs to define the optimal position of a Flying Gateway (FGW), aiming at accommodating the traffic demand. However, the GWP algorithm follows an iterative approach, which may lead to long execution times to achieve the optimal solution. This may be a problem when using Edge Computing, especially for large and highly dynamic flying networks where the dynamic traffic demand and the movements of the users may induce frequent changes in the placement of the FAPs.\looseness=-1

The main contribution of this paper is a fast gateway placement (F-GWP) algorithm for flying networks with controlled topology. Built upon the GWP algorithm, F-GWP is able to determine the optimal position of a FGW, in order to establish backhaul communications paths between the FAPs and the FGW able to accommodate the traffic demand of the users on the ground, but with execution times two orders of magnitude lower than GWP. This makes F-GWP especially suitable for highly dynamic flying networks using resource-constrained Edge Computing.\looseness=-1

The rest of this paper is organized as follows. 
\cref{sec:soa} presents the state of the art on UAV placement algorithms for flying networks. \cref{sec:system_model} defines the system model. 
\cref{sec:problem_formulation} formulates the problem addressed by this paper. 
\cref{sec:proposed_algorithm} presents the F-GWP algorithm. 
\cref{sec:performance_evaluation} describes the evaluation of the F-GWP algorithm, including the networking scenarios employed, the performance metrics adopted, the performance results achieved, and their discussion.
Finally,
\cref{sec:conclusions} presents the main conclusions and directions for future work.\looseness=-1

\section{State of the Art~\label{sec:soa}}
Several placement algorithms have been proposed in the literature to determine the optimal positions of UAVs forming flying networks. They are focused on addressing different challenges, including power consumption of the communications nodes and QoS guarantees \cite{kang2020}. Typically, the algorithms rely on deterministic approaches. A reference example is presented in \cite{lyu2016}, where the authors propose an algorithm that initially places the UAVs in the boundaries of a specific area composed of a set of users and afterwards follows an inward spiral until all the users are covered. Similar algorithms have been proposed in \cite{Mozaffari2015} and \cite{alzenad2017}, where the objective is to maximize the radio coverage and minimize the required transmission power. These algorithms are computationally efficient; however, they typically converge to solutions that are not optimal from the network performance point of view, since the traffic demand of the users is not considered. The location of the users to be covered is also commonly considered to place the UAVs. In \cite{yu2018}, an algorithm that places the UAVs vertically projected to the centroids of clusters of users is proposed, aiming at minimizing the required transmission power. An algorithm for placing UAVs that use directional antennas, taking into account a desired coverage area, is presented in \cite{mozaffari2016_EfficientDeployment}. It results in non-overlapping coverage areas formed by UAVs employing the minimum transmission power. Nevertheless, these algorithms are not traffic-aware, thus not providing QoS guarantees.\looseness=-1

Stochastic UAV placement approaches have also been proposed. A fast algorithm for placing a UAV providing connectivity to users with different QoS requirements is proposed in \cite{chen2018}. The authors formulated the problem as a Mixed Integer Second Order Cone Problem and studied the relationship between the coverage radius and the height of the UAV. Moreover, they employed the Standard Genetic Algorithm and Multi-Population Genetic Algorithm, in order to maximize the number of covered users, by employing random UAV positions that follow natural selection processes, including selection, crossover, and mutation. In \cite{na2019}, a UAV placement algorithm based on Particle Swarm Optimization (PSO) is proposed. It aims at determining the positions of the UAVs over time such that the connectivity between them is ensured and the aggregate sensor data information in a wireless sensor network is maximized. For that purpose, the information about the positions of the communications nodes is considered to be known in advance. However, these algorithms rely on modeling each UAV as an individual particle and adjusting its movement according to a specific utility function, which may result in long execution times to achieve the optimal solution.\looseness=-1

Overall, the state of the art UAV placement algorithms do not ensure QoS guarantees and low execution times simultaneously.\looseness=-1

\section{System Model~\label{sec:system_model}}
We assume a flying network organized into a two-tier architecture and composed of two types of UAVs, as depicted in Fig.~\ref{fig:music_festival_scenario}: 1) the FAPs, which are in charge of providing connectivity to the users on the ground, by enabling high-capacity small cells; and 2) the FGW, which forwards the traffic to/from the Internet. The FAPs are dynamically placed according to the traffic demand of the users on the ground by means of a state of the art FAP placement algorithm, such as the one proposed in \cite{Almeida2018}. For that purpose, the FAPs periodically take a snapshot of the network, including the positions of the users and the amount of traffic they are offering, and send this information to an Edge node, where the FAP placement algorithm is running; the Edge node may be one of the UAVs that compose the flying network or some node on the ground connected to the flying network. In turn, the optimal FGW placement is determined taking into account the future positions of the FAPs, which are defined by the FAP placement algorithm, as well as their traffic demand. The traffic demand of each FAP corresponds to the aggregate traffic exchanged with the users connected to it.\looseness=-1  

The interaction between the FAP and FGW placement algorithms takes advantage of the holistic and centralized view provided by the Edge node, where both algorithms run in parallel. This allows the use of the information provided by the FAP placement algorithm to calculate in advance the optimal FGW placement. Finally, the Edge node sends the updated positions to the FAPs and the FGW, which position themselves accordingly, enabling a seamless topology reconfiguration and maximizing the QoS offered by the flying network.\looseness=-1 

\section{Problem Formulation~\label{sec:problem_formulation}}
In order to model the wireless links, the Free-space Path Loss model expressed in \cref{eq:free-space-path-loss} in logarithmic scale is employed, since the wireless links between UAVs flying dozen of meters above the ground are characterized by a strong Line of Sight (LoS) component. In \cref{eq:free-space-path-loss}, $P_{R}$ represents the received power at the FGW, $P_T$ is the transmission power, $d$ represents the Euclidean distance between the FAPs and the FGW, $f$ stands for the carrier frequency of the wireless links established between the FAPs and the FGW, and $c$ is the speed of light in vacuum.\looseness=-1 

\begin{equation}
    \begin{aligned}
	P_R = P_T - 20 \smash{\times} \smash{\log}_{10}(d) - 20 \smash{\times} \smash{\log}_{10}(f) - 20 \smash{\times} \smash{\log}_{10} \bigl(\begin{smallmatrix}
	\frac{4 \smash{\times} \pi}{c}
	\end{smallmatrix}\bigr)
	\end{aligned}
	\label{eq:free-space-path-loss}
\end{equation} 

We assume that the maximum capacity of the wireless links results from the data rate associated to the Modulation and Coding Scheme (MCS) index selected by the communications nodes. The selection of a given MCS index requires a minimum Signal to Noise Ratio (SNR), $SNR = P_R - P_{Noise}$, considering a constant noise power $P_{Noise}$, according to the rationale proposed in \cite{Coelho2019}. The wireless medium is shared and we assume that all UAVs composing the flying network can listen to any other UAV. In order to avoid collisions of network packets, the Carrier Sense Multiple Access with Collision Avoidance (CSMA/CA) mechanism is assumed to be employed, which enables transmissions only when the wireless channel is sensed to be idle.\looseness=-1

The flying network is represented as a directed graph $G = (U, L)$, in which $U = \{UAV_1, ..., UAV_N\}$ is the set of UAVs $i$ placed at $P_i = (x_i, y_i, z_i)$, in a venue with dimensions $X \times Y \times Z$, and $L \subseteq U \times U$ represents the set of wireless links between $UAV_i$ and $UAV_j$, where $i, j \in U$. The flying network active topology is composed of single-hop paths; it is represented by a tree $T(U, L_T)$, rooted at $UAV_N$, which constitutes a subgraph of $G$, where $L_T \subseteq L$ is the set of wireless links established between $UAV_i$ and $UAV_N$, performing the role of FGW.\looseness=-1

Let us assume that $UAV_i, i \in \{1, ..., N-1\}$, acts as FAP and transmits a traffic flow $F_{i, N}$ with bitrate $T_i$ bit/s towards $UAV_N$. The maximum channel capacity is equal to $C^{MAX}$ bit/s, which is assumed to be equal to the data rate of the maximum MCS index of the wireless technology used. The maximum transmission power allowed for the wireless technology used is defined as $P^{MAX}$. The capacity of the bidirectional wireless link established between $UAV_i$ and $UAV_N$ at time $t_k$, where $t_k = k \cdot \Delta t, k \in \mathbb{N}_0, \Delta t \in {\rm I\!R}$ is equal to $C_{i, N}(t_k)$. $\Delta t$ is the update period imposed by the FAP placement algorithm, which considers a trade-off between the stability of the flying network and the time it takes to react to changes in traffic demand. Considering $N-1$ UAVs transmitting a traffic flow $F_{i,N}$ with bitrate $T_i(t_k)$ bit/s towards $UAV_N$, the problem consists of determining the position of $UAV_N$, $P_N = (x_N, y_N, z_N)$, such that the transmission power $P_T$ is minimized, while ensuring a minimum $SNR_{i}$ that guarantees $C_{i,N}(t_k)$ is high enough to accommodate $T_i(t_k)$ bit/s. The objective function is presented in \cref{eq:objective-function}.\looseness=-1

\begingroup
    \allowdisplaybreaks
    \begin{subequations}
    	\begin{alignat}{10}
    		  & \!\underset{(x_N, y_N, z_N)}{\text{minimize}} &  & \qquad P_T\label{eq:objective-function} \\
    		  & \text{subject to:} \notag\\
    		  &       &        &(N, i), (i, N) \in L_T, i \in \{1, ..., N-1\} \label{eq:constraint1} \\ 
    		  &       &        & 0 \leq P_T \leq P^{MAX} \label{eq:constraint2} \\
    		  &       &        & -\frac{X}{2} \leq x_i \leq \frac{X}{2}, i \in \{1,...,N\}\label{eq:constraint3} \\
    		  &       &        & -\frac{Y}{2} \leq y_i \leq \frac{Y}{2}, i \in \{1,...,N\}\label{eq:constraint4} \\
    		  &       &        & -\frac{Z}{2} \leq z_i \leq \frac{Z}{2}, i \in \{1,...,N\}\label{eq:constraint5} \\
    		  &       &        & (x_N, y_N, z_N) \neq (x_i, y_i, z_i), i \in\{1\smash {,}...\smash {,}N\smash {-}1\}\label{eq:constraint6} \\
    		  &       &        & (x_N - x_i)^2 + (y_N - y_i)^2 + (z_N -  z_i)^2 \notag\\
              &       &        & \leq (10^{\frac{K + P_T - SNR_i}{20}})^2, i \in \{1, ..., N-1\} \label{eq:constraint7}\\
              &       &        & K = - 20\log_{10}(f) -20\log_{10}\left(\frac{4\pi}{c}\right)  \smash {-} P_{Noise} \label{eq:constraint8}\\
              &       &        & \sum_{i=1}^{N-1}C_{i,N}(t_k) \leq C^{MAX}, i \in\{1,...,N-1\}\label{eq:constraint9} \\
              &       &        &  0 < T_i(t_k) \leq C_{i,N}(t_k), i \in \{1,...,N-1\}\label{eq:constraint10}
    	\end{alignat}
    \end{subequations}
\endgroup

The factors that influence the calculation of $P_N = (x_N, y_N, z_N)$ and $P_T$ include a) the capacity of the wireless links between $UAV_i$ and $UAV_N$ at each instant $t_k$, which is affected by the number of communications nodes that use the same wireless channel and the behavior of the medium access protocol, and b) the interference between the communications nodes. Since these factors are difficult to model in order to solve the problem analytically, in this paper we propose a heuristic algorithm to achieve a solution.\looseness=-1

\section{Fast Gateway Placement Algorithm ~\label{sec:proposed_algorithm}}
In this section, the proposed algorithm is presented, including its rationale and implementation.

\subsection{Rationale}\label{subsec:rationale}
The fast gateway placement (F-GWP) algorithm for flying networks was built upon the GWP algorithm, originally proposed in \cite{Coelho2019}. Similarly to GWP, the F-GWP algorithm relies on the holistic and centralized view provided by the Edge node, namely the future positions of the FAPs, defined by a state of the art FAP placement algorithm (e.g., the one proposed in \cite{Almeida2018}), and the traffic demand of the FAPs, in order to determine the optimal position of the FGW. The F-GWP algorithm aims at computing the minimum transmission power for the FAPs and the FGW that allows establishing wireless links between $UAV_i$ and $UAV_N$ with high enough capacity to accommodate the traffic demand $T_i$. For that purpose, the F-GWP algorithm initially determines the minimum $SNR_i$ that enables the selection of an MCS index, $MCS_i$, equivalent to a data rate capable of accommodating $T_i$ bit/s \cite{MCSIndex14:online}. Since the medium is assumed to be shared by multiple FAPs, the fair share of the wireless channel capacity is considered for defining the traffic demand $T_i$, which results from the data rate associated to $MCS_i$ over the number of FAPs making part of the flying network, as explained in \cite{Coelho2019}.

\begin{figure}
	\centering
	\includegraphics[width=0.75\linewidth]{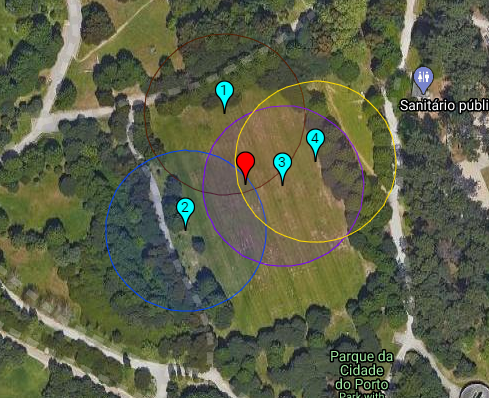}
	\caption{Aerial view of the intersection of the FAPs' transmission ranges and F-GWP solution (unnumbered red marker) at Porto City Park \cite{parqueporto:online}.}
	\label{fig:aerial_intersection}
\end{figure}

The minimum $SNR_i$ that enables the selection of $MCS_i$ imposes a minimum received power $P_{R_i}$. Taking into account the transmission power of the FAPs, $P_T$, then the maximum admissible distance $d_i$ between $UAV_i$ and $UAV_N$ can be calculated by means of the Free-space Path Loss model -- cf. \cref{eq:free-space-path-loss}. In the three-dimensional space, $d_i$ represents the radius of the sphere centered at $UAV_i$, corresponding  to its transmission range, within which $UAV_N$ should be placed.
Considering $N-1$ UAVs acting as FAPs, $UAV_N$ should be placed in the volume defined by the intersection of the spheres centered at each $UAV_i$, as depicted in \cref{fig:aerial_intersection}. 

The F-GWP algorithm is in charge of determining the minimum transmission power $P_T$ that enables the intersection of all the spheres. While the GWP algorithm determines the value of $P_T$ following an iterative approach, successively increasing $P_T$ by 1 dBm from 0 dBm onwards until the intersection between the spheres occurs, the F-GWP algorithm solves an optimization problem that aims at minimizing $P_T$, assumed to be equal for all FAPs and the FGW. For this reason, while the GWP algorithm may lead to long execution times to achieve the optimal solution, the F-GWP algorithm is able to determine the optimal FGW placement and the minimum transmission power of the FAPs and the FGW in shorter time. This makes it especially suitable for highly dynamic flying networks using resource-constrained Edge Computing.

\subsection{Algorithm Implementation~\label{sec:algo_impl}}
To implement the F-GWP algorithm, we took advantage of the package \emph{SciPy} \cite{scipy:online} for \emph{Python} and its companion tool \emph{optimize} \cite{optimize:online}. This allows to solve the problem formulated in \cref{sec:problem_formulation} employing an iterative method. Using this framework, we first set the objective function and the problem constraints. Then, we defined the initial positions of the UAVs and set the initial transmission power to \SI{1}{\deci\belmilliwatt}, which were considered by the solver to achieve the solution. Sequential Least SQuares Programming \cite{iserles1995acta} was used, which is an iterative method targeted for constrained non-linear optimization.

Regardless of the direction of the traffic -- downlink or uplink -- the output of the F-GWP algorithm is the same; all UAVs are configured with the same minimum transmission power and the wireless channel is assumed to be symmetric. In practice, the wireless channel is asymmetric. Yet, the algorithm will still work, if in the computation of the maximum distance $d_i$ the loss for the worst direction (downlink or uplink) is considered; this basically implies the sum of the additional loss to the value obtained using the Free-space Path Loss model (cf. \cref{subsec:rationale}). In this way, it is possible to use the F-GWP algorithm in emerging networking scenarios where applications generating symmetric traffic predominate~\cite{elshaer2014}, including social networks, video streaming, and online gaming.

\section{Performance Evaluation~\label{sec:performance_evaluation}}
The performance evaluation of the F-GWP algorithm is presented in this section, including the networking scenarios employed, the performance metrics adopted, and the results obtained and their discussion.

\subsection{Networking Scenarios~\label{sec:networking_scenarios}}
In order to evaluate the performance of the F-GWP algorithm, different networking scenarios were generated using the IEEE 802.11ac standard \cite{802.11acStandard}, taking into account an area \SI{15}{\meter} long and wide, and \SI{20}{\meter} high. Flying networks with 2, 4, 6, 8, 10, 15, and 20 FAPs were considered. For each number of FAPs, 10 random networks were generated. For each FAP, the minimum $SNR_i$ was defined between \SI{11}{\deci\bel} and \SI{40}{\deci\bel}, which enables data rates from 58.5 Mbit/s to 780 Mbit/s, respectively, the channel bandwidth was set to \SI{160}{\mega\hertz}, and the Guard Interval (GI) was set to \SI{800}{\nano\second}. The minimum $SNR_i$ was defined taking into account the relation between the SNR and the data rates for the IEEE 802.11ac MCS indexes \cite{MCSIndex14:online}, according to the rationale proposed in \cite{Coelho2019}.

\subsection{Evaluation Metrics~\label{sec:performance_metrics}}
The metrics used to compare the computational performance of the GWP and F-GWP algorithms were 1) the \textbf{execution time} (in seconds) required to obtain the solution -- i.e., the FGW position and the transmission power $P_T$ -- and 2) the \textbf{memory used} by each algorithm (in megabytes, \SI{}{\mega\byte}).

In order to evaluate the network performance achieved with the F-GWP algorithm in comparison with the GWP algorithm, the following networking metrics were considered: 

\begin{itemize}
	\item \textbf{Aggregate throughput} -- the mean number of bits received per second by $UAV_N$ (FGW);
	\item \textbf{Delay} -- the mean time taken by the packets to reach the sink application running at $UAV_N$, since the instant of time they were generated by the source application running at each FAP.
\end{itemize} 

\subsection{Computational Performance Results~\label{sec:computational_performance_results}} 

\begin{figure}
	\centering
	\subfloat[Execution time for different number of FAPs.]{
		\includegraphics[width=0.8\linewidth]{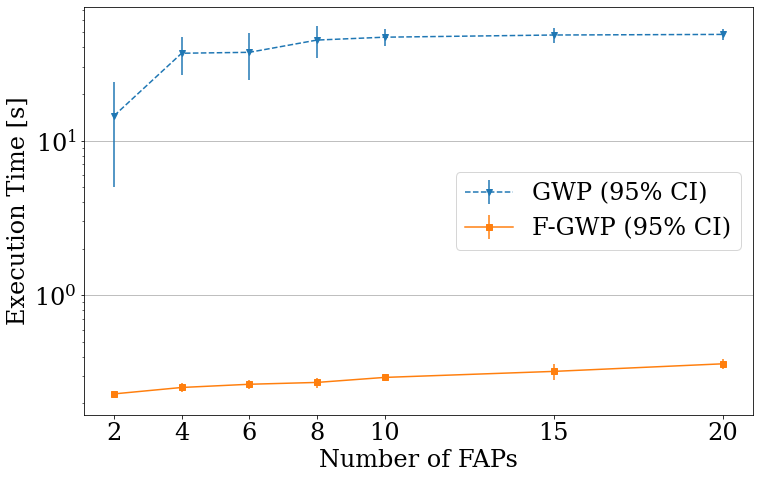}
		\label{fig:time-performance}}
	\hfill
	\subfloat[Memory usage for different number of FAPs.]{
		\includegraphics[width=0.8\linewidth]{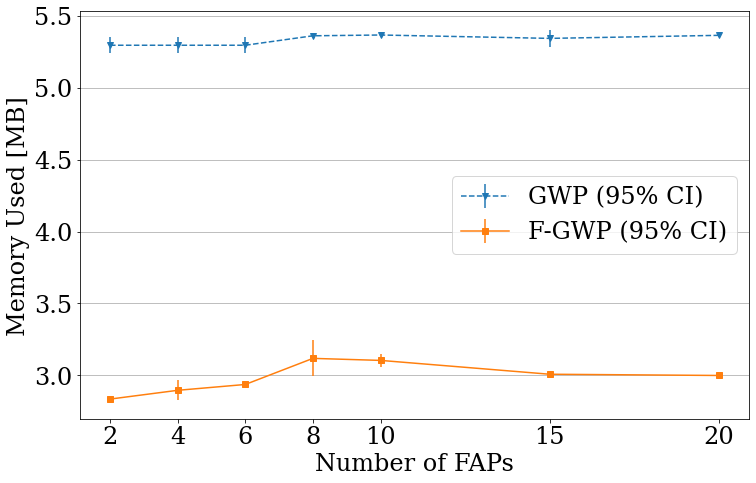}
		\label{fig:memory-performance}}
	\captionof{figure}{Computational performance results, in terms of execution time and memory usage, considering different number of FAPs. The F-GWP algorithm (square markers) outperforms the GWP algorithm (round markers) in both metrics.}
	\label{fig:performance_graph}
\end{figure}

The computational performance evaluation was carried out on a \emph{Google Colaboratory} \cite{colab:online} Jupyter Notebook, in a hosted runtime with an Intel® Xeon® processor running at \SI{2.20}{\giga\hertz} (2 threads) and 13 GB of RAM.
As we can observe in the plot of \cref{fig:time-performance}, the F-GWP algorithm is around 100 times faster than the GWP algorithm. For both small and large networks, the execution time is in the order of tens of milliseconds. The increase is linear with the number of FAPs making part of the flying network. The F-GWP algorithm, besides being faster than the GWP algorithm, proves to be much more stable as it has very low variability in almost all scenarios, as shown by the very short 95\% confidence intervals.

Regarding memory usage, which is often neglected, we observe in \cref{fig:memory-performance} that F-GWP outperforms GWP using about \SI{3}{\mega\byte} in the tested scenarios, when compared to the 5+ MB used by GWP. Both algorithms are stable, with low variability and slow increase as the network size increases. Memory usage was determined by measuring how much memory the \emph{Python} interpreter was using immediately before and after running the GWP and F-GWP algorithms. The \emph{Python}'s garbage collector, responsible for memory management, was used in automatic mode, which may justify the peak around 8 FAPs that is observable in \cref{fig:memory-performance}.

\subsection{Network Simulation Setup \label{sec:simulation-setup}}
In order to evaluate the network performance when the F-GWP algorithm is employed and compare it against the GWP algorithm, the ns-3 simulator \cite{ns3Simulator} was used. The FAPs and the FGW were carrying a Network Interface Card (NIC) in Ad Hoc mode, which was set to use the IEEE 802.11ac technology in channel 50, \SI{160}{\mega\hertz} channel bandwidth, and \SI{800}{\nano\second} Guard Interval. The wireless link established between each FAP and the FGW was using a single spatial stream. The FAPs were generating UDP Poisson traffic towards the FGW. The data rate was defined by the \emph{IdealWifiManager} mechanism.

\subsection{Network Performance Results\label{sec:network-performance-results}}
The network performance results were obtained by means of ns-3 simulations, which consisted of 20 runs under the same exact networking scenarios employed in the computational performance evaluation presented in \cref{sec:computational_performance_results}, considering \emph{RngSeed = 20} and \emph{RngRun = \{1, ..., 20\}}. The results are expressed as mean values and they are presented by using 1) the Cumulative Distribution Function (CDF) for the delay that represents the percentage of samples for which the mean packet delay was lower than or equal to $x$, and 2) the complementary CDF (CCDF) for the aggregate throughput that represents the percentage of samples for which the aggregate throughput was higher than $x$.

\begin{figure}
	\centering
	\subfloat[Throughput CCDF.]{
		\includegraphics[width=1\linewidth]{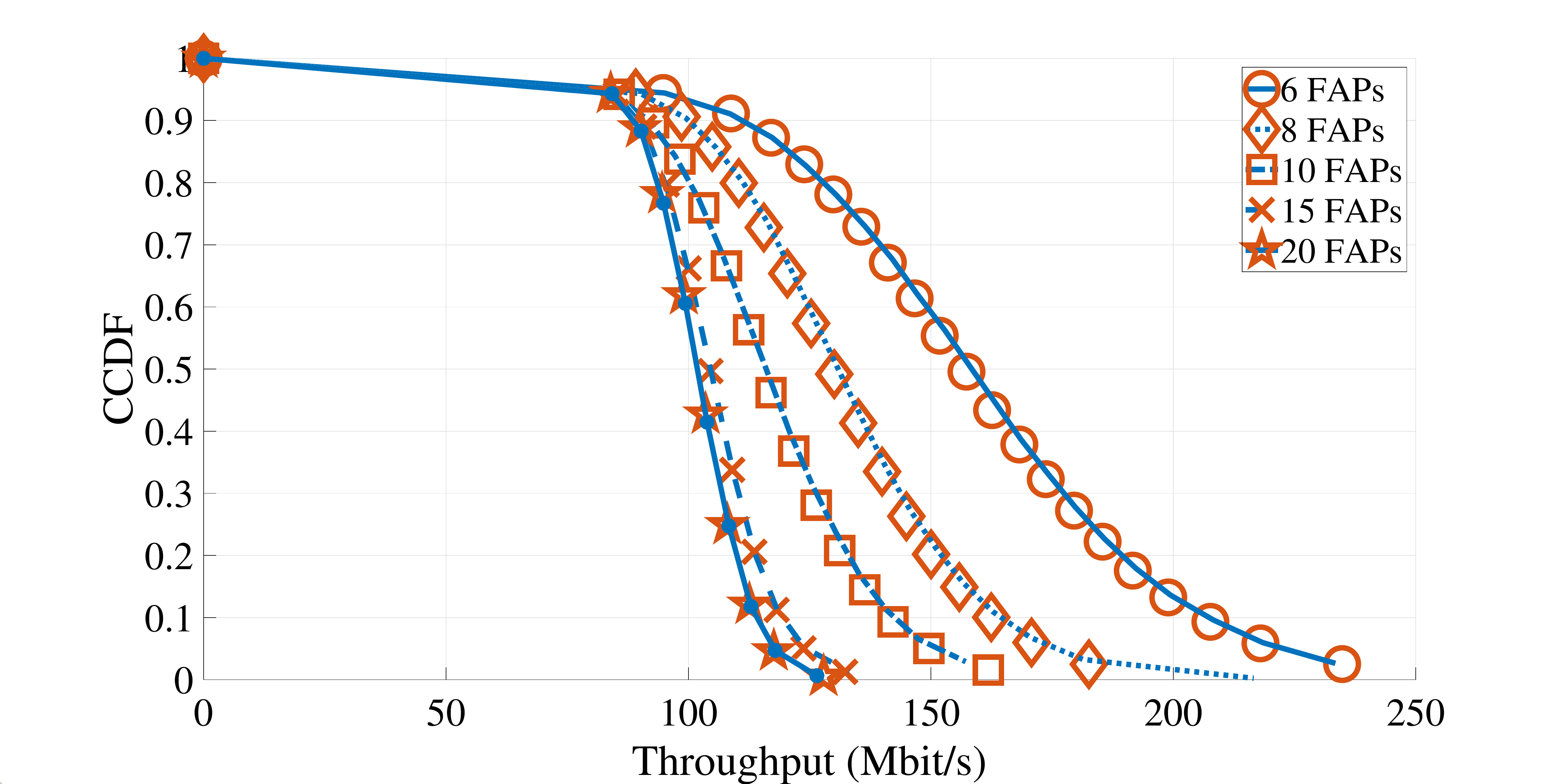}
		\label{fig:throughput-ccdf}}
	\hfill
	\subfloat[Delay CDF.]{
		\includegraphics[width=1\linewidth]{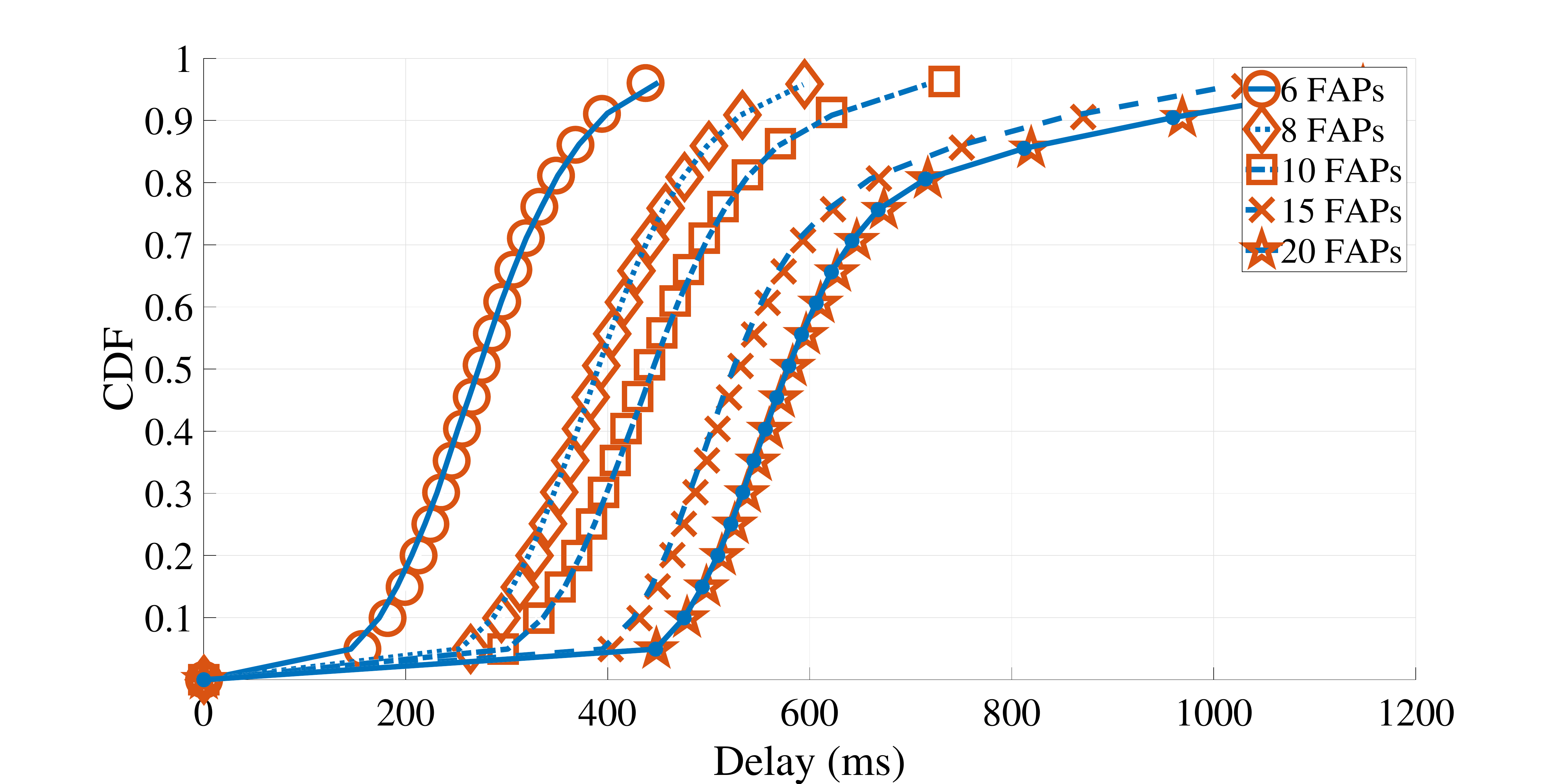}
		\label{fig:delay-cdf}}
	\caption{Network performance results, in terms of aggregate throughput and mean delay, considering different number of FAPs. The F-GWP algorithm (markers) enables the same network performance as the GWP algorithm (lines).}
	\label{fig:network-performance-results}
\end{figure}

The obtained results, measured in the FGW, are presented in \cref{fig:network-performance-results}. They represent the network performance when the FGW is placed in the positions defined by 1) the F-GWP algorithm (markers), and 2) the GWP algorithm (lines), considering the transmission power $P_T$ calculated by each algorithm. The plots clearly show the F-GWP algorithm achieves the same network performance as the GWP algorithm, for which the network performance evaluation presented in \cite{Coelho2019} allows to conclude significant gains regarding aggregate throughput and delay, while enabling decisions up to 100 times faster (cf. \cref{sec:computational_performance_results}).

\subsection{Discussion~\label{sec:discussion}}

There is a significant difference between the execution times of both algorithms for the same network performance, namely when the number of FAPs in the flying network increases. This happens because, in the GWP algorithm, for each additional FAP an additional equation has to be solved as part of the system of equations used to determine the position of the FGW \cite{Coelho2019}. While the execution times are higher even for a low number of FAPs, adding more equations as the number of FAPs increases does not have a major impact on the computational performance, as it can be observed in \cref{fig:performance_graph}. 

On the other hand, the F-GWP algorithm uses the equations as constraints of the optimization problem, which leads to lower execution times in general. Yet, a higher number of FAPs in the flying network has an increased impact, since it implies more constraints to be taken into consideration, thus increasing the execution time. Since the size of the flying networks that are expected to be used in practice is limited, we do not expect to reach a point where the F-GWP algorithm does not prevail over the GWP algorithm.

The same behavior is observed when comparing the memory usage of both algorithms, for which the difference between them is another aspect in favor of the F-GWP algorithm. This improvement is relevant if we consider the F-GWP algorithm will run in an Edge node, which is resource-constrained.

The fact that we used \emph{Python} and third-party packages (cf. \cref{sec:algo_impl}) to implement both algorithms leaves room for improvement. Some other low-level languages, such as \emph{C} and \emph{Java}, which can lead to better computational performance for both algorithms, are worthy to be explored.

\section{Conclusions~\label{sec:conclusions}}
This paper proposed a fast gateway placement algorithm for flying networks, named F-GWP, which is able to determine the optimal position of a FGW in shorter time. The experimental results show the F-GWP algorithm has significant gains from the execution time point of view, enabling decisions up to 100 times faster than the state of the art GWP algorithm, while ensuring backhaul communications paths with high enough capacity to accommodate the traffic demand of the users. This makes F-GWP especially suitable for highly dynamic flying networks using resource-constrained Edge Computing. As future work, we aim at exploring the F-GWP algorithm in multi-FGW and multi-hop flying networks composed of UAV relays forwarding traffic between the FAPs and the FGWs.

\section*{Acknowledgments}
This work is co-financed by National Funds through the Portuguese funding agency, FCT -- Fundação para a Ciência e a Tecnologia, within project UIDB/50014/2020 and by the European Regional Development Fund (FEDER), through the Regional Operational Programme of Lisbon (POR LISBOA 2020) and the Competitiveness and Internationalization Operational Programme (COMPETE 2020) of the Portugal 2020 framework [Project 5G with Nr. 024539 (POCI-01-0247-FEDER-024539)]. The third author also thanks the funding from FCT under the PhD grant SFRH/BD/137255/2018.

\bibliographystyle{IEEEtran}
\bibliography{IEEEabrv,References}

\end{document}